# Declining trends of healthy life years expectancy (HLYE) in Europe.

Ugo Bardi[*] and Virginia Perini

Dipartimento di Scienze della Terra – Università di Firenze

c/o Dipartimento di Chimica - Polo scientifico di Sesto Fiorentino

50019 Sesto Fiorentino, Fi, Italy

* Author to whom correspondence should be addressed; ugo.bardi@unifi.it; Tel.: +39 055-457-3118; Fax: +39-055-457-3120.

**Abstract**

We examine the trends in Healthy Life Years Expectancy (HLYE) at birth during the past few decades in Europe. We observe that several European countries show a significant drop in HLYE at birth starting from 2003; interrupting what had been a previous continuous increase. We discuss the possible causes of this drop, including a possible correlation with climate change in terms of the heat wave experienced in Europe in 2003. It is not possible, at present, to propose a single explanation for this phenomenon, however the trend is worrisome and its causes should be investigated further.

## 1. Introduction

It is often stated that environmental factors such as climate change and pollution are not so important in affecting human health because life expectancy (LE) continues to increase, at least in developed countries. However, life expectancy is not the only indicator of people's health and quality of life. The Healthy Life Years Expectancy (HLYE) (also called Disability-Free Life Expectancy, DFLE) measures the number of expected disability-free years. HLYE is mainly affected by chronic diseases, which have become the main cause of death in developed countries and it may be sensitive to factors such as ambient temperatures, in turn affected by climate change.

Our purpose with the present study was to examine the HLYE trends during the past few decades and to see if similar effects can be observed. For this aim we compared the parameters of LE and HLYE at birth using data relative to the 27 states of the European union, focusing in particular on Italy. We found a considerable drop in the HLYE parameter of several European countries that occurred starting with 2003.

Several factors can be considered as the cause of the observed drop in HLYE and it is possible to propose a correlation with the 2003 heat wave experienced in Europe as it has already been observed that life expectancy can be negatively affected by climatic factors such as the European heat wave of 2003 [1]. Nevertheless, the correlation holds only for some European countries and hence further studies will be necessary to understand the causes of this phenomenon. Even so, the authors of the present study felt that it was important to diffuse these data to stress the importance of HLYE as a health indicator and of the fact that an increase in life expectancy is not necessarily an indication of better living conditions of the population.

## 2. Method

Life expectancy (LE) is a parameter that defines the expected average number of years of life remaining at a given age. This indicator is normally used to measure the average duration of life and to compare different groups of people. An increase in life expectancy at birth is normally considered an indication of better health conditions of a population.

In developed countries, the LE parameter has been continuously increasing throughout the 20$^{th}$ century and the first decade of the 21$^{st}$. This increase is normally related to the reduction of mortality from infectious diseases. However, the fact of living longer does not rule out a lowered quality of life, considering that today the cause of death in developed countries is mainly related to chronic illnesses. The parameter defined as Healthy Life Years Expectancy (HLYE) measures the number of years that a person is expected to live in a healthy condition, if subjected throughout the rest of his or her life to the current mortality conditions. A healthy condition is defined by the absence of limitations in functioning/disability.

The idea of healthy life expectancy has been advanced for the first time by [2] and a first method of calculation was proposed by Sullivan [3]. Today, the Sullivan's method is the most commonly used method for calculating health expectancies [4]. It combines mortality data from population life tables and age specific prevalence of the population in healthy or unhealthy conditions obtained from surveys [5]. The European Community Household Panel (ECHP) provides a longitudinal, multi-subject survey covering many aspects of daily life. Although data from the ECHP should theoretically provide concordant data, changes over time and differences between countries in the survey design and question wording have required some adjustments to be made before calculations [6]. Standardized translations of the questionnaire have been used; nevertheless it is likely that linguistic or cultural differences, as well as changes in the wording of questions that were implemented in 2004, have influenced the way the respondents indicate a longstanding health problem or disability and their way of communicating the types of restrictions caused by this problem [7].

## 3. Data analysis

Starting from the case of Italy, we report in fig. 1 data for both LE and HLYE. We can see that from 1995 to 2011 life expectancy (LE) at birth in Italy increased from 75 to 80,1 for men and, in the same years, from 81,8 to 85,3 for women. In the same period, the Healthy Life Years Expectancy (HLYE) at birth showed a completely different trend. From 1995 to 2003 the HLYE values in Italy rose up for both genders: from 66,7 to 70,9 years for men and from 70 to 74,4 years for women. However, after 2003 HLYE rapidly decreases, to stabilize in 2008 at a much lower value (around 62 years) for both genders, although the data for 2010 show a temporary significant increase. According to these rates, Italian women seem to live at least 5 years longer than men, but start suffering of disabilities at about the same age.

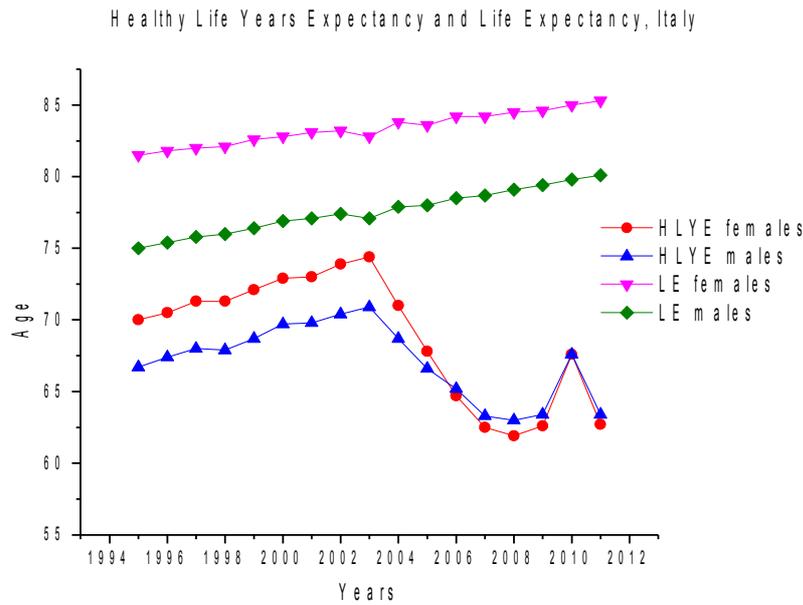

**Fig. 1** Healthy Life Years Expectancy and Life Expectancy at birth in Italy – males and females – Years 1995-2011.

Italy is the country for which the 2003 drop in HLYE is most significant, but it is by no means an isolated case. From the Eurostat data ( [8] and [9] ) the 2003 drop occurs also in other European Union Member States: Belgium, Germany, Ireland, Greece, Spain, Austria, Portugal, Finland and Sweden (Fig. 2).

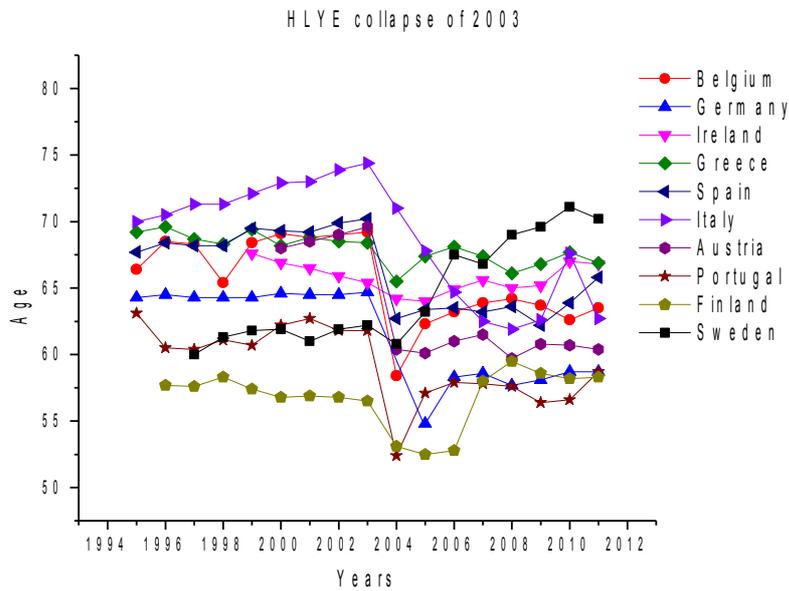

**Fig. 2** Healthy Life Years Expectancy at birth (females) in Belgium, Germany, Ireland, Greece, Spain, Italy, Austria, Portugal, Finland and Sweden – Years 1995-2011.

The main results of our analysis are summarized in the following tables and figures.

**Table 1**

| Countries showing declining HLYE after 2003 |
|---|
| Belgium |
| Germany |
| Ireland |
| Greece |
| Spain |
| Italy |
| Austria |
| Portugal |
| Finland |
| Sweden |

**Table 2**

| Countries with a non-declining HLYE after 2003 |
|---|
| Denmark |
| France |
| Netherlands |
| UK |

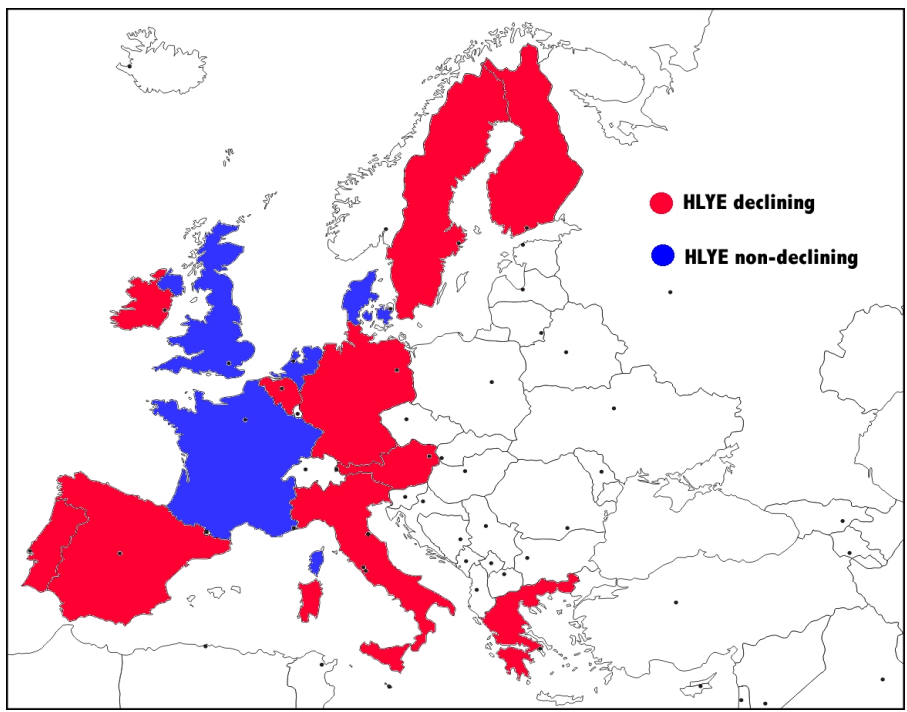

**Fig. 3** The countries with a declining HLYE after 2003 are in red; the countries with a non-declining HLYE after 2003 are in blue.

In the following, we show the data for France, a country that showed no evidence of a decline in HLYE in 2003, but saw a stasis and a slow decline starting from approximately 2006.

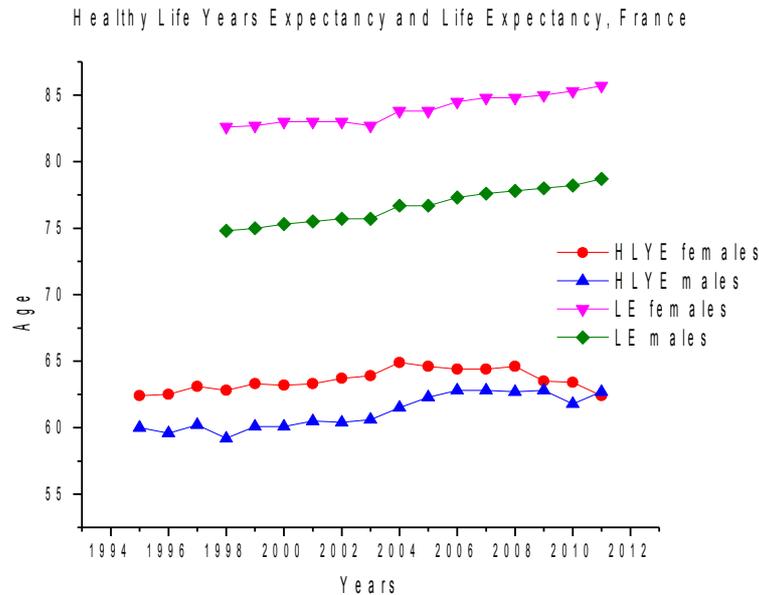

**Fig. 4** Healthy Life Years Expectancy and Life Expectancy at birth in France – males and females – Years 1995-2011.

## 4. Discussion

The data on Healthy Life Years Expectancy is obtained by direct questioning people and such a determination is obviously more uncertain and subjective than the determination of the event of one's death. However, what we are measuring here is the variation, rather than the absolute value, of the Healthy Life Years Expectancy. Since the change observed is considerable for several countries, being in a range of several years of difference, we are clearly seeing a significant effect related to people's health or, at least, of people's perception of their own health, or different ways in the way the question was understood depending on how it was worded in different languages. All or each of these effects in themselves may be indicating changes in the health status of European citizens.

Unfortunately, it is not possible to find a common denominator for the data reported in the previous section. While several countries show a drop in HLYE in 2003, others don't. In France, we see a small drop in the overall life expectancy for both males and females, but no effect – or perhaps a slight increase, in the HLYE. Nevertheless, France does sees a stasis and a slow reduction of HLYE starting approximately with 2006. In UK, we also see a small drop in life expectancy in 2004, but a remarkable increase (the sole case on record) in HLYE in the same year. The data do not show an evident correlation of the effect with the latitudes of the countries affected, not to other parameters relative, for instance, to income per capita or to the quality of the health care system. For example, countries with a high per capita income, such as Germany and Sweden, are affected by the 2003 HLYE drop just as countries with a lower per capita income, such as Italy, Greece and Spain.

We may nevertheless propose a tentative correlation of the HLYE trends with the heat wave experienced in Europe during the Summer of 2003. It is well known that this heat wave generated increased mortality. In August 2003 alone, nearly 45,000 additional deaths were recorded in the twelve countries examined by the "Report on excess mortality in Europe during summer 2003"[10] including 15,251 in France (+37%), 9,713 in Italy (+21.8%), 7,295 in Germany (+11%), 6,461 in Spain (+22.9%) and 1,987 in England and Wales (+4.9%). In comparison with the 1998-2002 period, more than 80,000 additional deaths were recorded in 2003 in the twelve countries concerned by excess mortality. In Italy, compared with the same period of 2002, the heat wave generated 3,134 additional deaths (from 20,564 in 2002 to 23,698 in 2003). The greatest increase was among the elderly: 2,876 deaths (92%) occurred among people aged 75 years and older [11]. Note how the increased mortality is visible as a small drop in the LE data for Italy (fig. 1) and for France (Fig. 4). This higher mortality is related to how high temperatures affect human health as the result of stress on the cardiovascular and respiratory systems [12] especially among elderly persons [11]. Note also that it is not just external temperature conditions that affect the health of elder people, but also the quality of the general living conditions and even the whole ecosystem [13]. Economic factors may also be relevant [14]. Chronic diseases have a strong impact on well-being as disability is one of their major consequences. Also factors such as atmospheric pollution (e.g. ozone, [15]) and/or water pollution and sanitary conditions [16] are affected by ambient temperatures and may have played a role.

Reasonably, actual deaths could not be the only result of these factors during the 2003 heat wave. So, it is at least possible to propose that the 2003 heat wave was among the factors involved in the decreased HLYE of several countries. There remains the problem to explain why the same effect was not observed in all countries examined and the fact that France, which heavily suffered the 2003 heat wave, does not show the same HLYE drop that we can observe in neighboring Italy. A further unexplained question is why in Italy the drop was so pronounced in comparison to other European countries. Such a large drop in the Italian HLYE might have caused a detectable increase in the national health costs. The available data do indicate a significant increase in these costs in the period 2003-2004 in Italy[17], however it is not possible at present to propose a significant correlation. In the end, a completely satisfactory explanation for the observed trends does not exist, but the possibility of an external (climatic?) effect in the HLYE drop cannot be ruled out.

## 5. Conclusions

This analysis of differences of LE and HLYE, at birth, shows that the average duration of life in Europe is increasing but also that, at the same time, the quality of life as measured by the HLYE parameter, has been declining in several countries after 2003, especially for the case of Italy. Also other countries, such as France, show a decline in HLYE starting at later dates. Apart from the differences among geographical areas, HLYE trends differ also between genders: women, who live longer, have generally a not better, and sometimes worse, health quality of life. As Europe was invested by a particularly strong heat wave on Summer 2003, a possible correlation between the HLYE breakdown and this unusual climate condition can be proposed, although it is clearly only a partial explanation. Further investigation elucidating these additional factors is required.

**Note**: after the first upload in ArXiv, on Nov 29[th] 2013, we discovered that the subject of declining healthy life expectancy had been noted in 2012 by Gennaro et al. [18] for the case of Italy alone. In a comment to this earlier paper, Piergentili suggested that the HLYE drop may be related to measurement uncertainties [19]

**Conflict of Interest**

The authors declare no conflict of interest.

**Acknowledgement**

The authors would like to thank Mr. Yvan Dutil for advice on improving this paper